\documentclass[jcp, preprint,onecolumn,noshowpacs]{revtex4}
\usepackage{epstopdf}
\usepackage{amssymb}
\usepackage{amsmath}
\usepackage{amsfonts}
\usepackage{graphicx}
\usepackage{color}
\usepackage{ulem}
\begin{document}

\title{THERMODYNAMICALLY CONSISTENT COARSE-GRAINING OF POLYMERS}
\author{M. G. Guenza\footnote{Author to whom correspondence should be addressed. Electronic mail: mguenza@uoregon.edu}}
\affiliation{Department of Chemistry and Biochemistry, and Institute of Theoretical Science, University of Oregon, Eugene, Oregon 97403}
\date{\today}

\begin{abstract}

\end{abstract}
\maketitle

Molecular dynamics simulations have become an essential tool in developing and testing theories in complex systems of either biological or synthetic origins. However, explicit atom molecular dynamics simulations are computationally costly, which limits their applicability to a restricted range of length and time-scales. A way to overcome this limitation is to design specific coarse-graining (CG) models. These models can be highly efficient with respect to atomistic descriptions because they represent the system at a lower resolution, thus greatly reducing the degrees of freedom that need to be sampled during the simulation.\cite{Hall,Dama,Martini,Arun,Kremer,depablo}

However, reducing the degrees of freedom comes with the consequence that some thermodynamic properties and the dynamical quantities are modified in a way that depend on the extent  of coarse-graining. This can be intuitively understood by considering that when a system is coarse-grained, a number of atomistic units are combined together into a new, effective, fictitious unit. Those CG units interact by means of an effective potential that is simplified and, in general, is smoother than the original atomistic potential. While the atomistic units spend time sampling a myriad of local configurations, the effective CG units can slide freely on a smoother free energy landscape. 

A coarse-graining procedure is formally equivalent to applying the Mori-Zwanzig projection operator technique to a macromolecular liquid, originally described at the level of the Hamiltonian by a Liouville equation. The projection simplifies the system to a number of coarse-grained units interacting through an effective free-energy-type potential in the reduced representation. The dynamics of the coarse-grained units is then described by a Langevin equation where dissipation emerges from the procedure.
Because the potential between units in the coarse-grained description is a free energy, it is parameter dependent and changes with the thermodynamic settings and the type of molecular system. We should expect a different potential to act between the coarse-grained units when the value of the temperature, the density, the molecular concentration, the number of monomers in the coarse-grained unit, or even the degree of polymerization of the molecule is varied. This is a problem for the proper design of a coarse-grained description if the potential is numerically derived, i.e. an analytical solution is not possible, leading to representability problems for the potential.

Most of the effective potentials in traditional Molecular Dynamics simulations, such as the Lennard-Jones potential or the united atom potential,\cite{UAmodel} are dependent on thermodynamic and molecular parameters in a trivial way. They depend on temperature and on the chemical identity of the particles involved but do not depend on other parameters, such as the number of atoms in the molecule or the molecular concentration; effective coarse-grained potentials usually present a more complex behavior. We argue that the capability for traditional potential of providing a reasonable estimate of the properties by molecular dynamics simulations, even in regions of the phase diagram where they perhaps should not be applied, is due to the small error thypical of CG models that have a very limited level of coarse-graining. 

Much of the difficulty in designing consistent coarse-graining approaches stems from the fact that it is not always clear how various many-body effects are incorporated into simple two-body interactions, and how errors in the numerical optimization of a pair potential are propagated to thermodynamic properties.\cite{Anthony2} The CG approach that we developed, and that we overview in this chapter, has the advantage of providing an analytical solution of the potential, structural properties, and thermodynamic properties as a function of the parameters in the system, which is represented with a variable level of resolution.\cite{Jaypaper}

In most CG models, which are numerically solved, the step of running detailed atomistic simulations is necessary to be able to parameterize the model. However, running detailed atomistic simulations in part defeats the purpose of developing accurate coarse-graining models, because a detailed atomistic description is already available for the system under study in the thermodynamic region of interest. The hope is that the models optimized against detailed atomistic simulations can perform well in regions of the phase space that is close to the point where the coarse-grained model has been optimized. This is a problem of transferability of the potential.

In general, numerically optimized CG potentials do not apply to regions in configurational space where they have not been optimized, and atomistic simulations should be performed at any condition of interest to test the applicability of the potential. However, after atomistic simulations are performed at any conditions of interest, it is not clear what would be the purpose of running also mesoscale simulations in the coarse-grained description.

One parameter that seems reasonable to vary is the time covered by the simulation, which in the CG description can be orders of magnitude longer than in the atomistic simulation. The underlying assumption is that the effective potential acts between sites, no matter the time covered by the simulations. However, in this case, increasing the time of the simulation could hardly bring new information, while in the opposite case the long-time prediction of the CG model could be incorrect. In general, it is difficult to know if a model parameterized in a short simulation contains information from all the relevant free energy barriers, some of which could come into play only at a timescale that exceeds the length of the initial atomistic simulation. The process of crossing those high energy barriers could exceed the timescale sampled in the atomistic simulation and the coarse-grained description derived from the atomistic simulations would likely predict incorrect long time behavior of the system.

In general, the key point in  analyzing the advantages and challenges of the different CG models is to understand which properties should be conserved, which properties should be modified in the process, and how they will be modified as a consequence of coarse-graining.
As a starting step, with the purpose of understanding the microscopic motivations of the observed thermodynamic and dynamical inconsistences, it is useful to consider an intuitive representation of the effects of coarse-graining. Here we look at a very simple model given by the comparison between the free energy landscape for the rotation of the dihedral angle in an ethane molecule in the coarse-grained and atomistic representations. In the atomistic, the energy presents three minima corresponding to one trans and the two gauche configurations. In the mesoscale representation of the united atom model, each methyl unit, $CH_3$, is  an effective CG sphere, the whole molecule is a dumb-bell, and the rotation of the two units with respect to each other is free, leading to a completely flat free energy landscape.

Several effects typical of coarse-grained descriptions arise from site-averaging, like in this example. First of all, the dynamics speeds up as the 
$CH_3$ units, which in the atomistic description employed a finite amount of time in crossing the configurational barriers that separate states of low energy, in the coarse-grained description are fully free to rotate. The speeding up of the dynamics gives the computational advantage to the coarse-grained description; the more extensive the coarse-graining the larger the gain in the computational time. The dynamic properties of the coarse-grained representations are, however, unrealistically accelerated: for example, the diffusion coefficient measured in mesoscale simulations can be orders of magnitude too high. One could be tempted to look into slowing down the coarse-grained model so that the simulation can reproduce the correct dynamics; however that choice would also slow down the computational time, loosing the gain of having a CG description. Different strategies to recover the correct dynamics in a fast computational framework have been implemented. For example, it is possible to directly modify the equations of motion for the CG units to reproduce the correct dynamics in a coarse-grained description, as it is done in Dissipative Particle Dynamics.\cite{DPD} Another strategy is to use reduced dynamical models such as a Generalized Langevin representation of a restricted number of  molecules moving slowly in the field of the fast-moving molecules (solvent or polymers). This strategy was applied in a model we developed to describe protein dynamics.\cite{Copperman,Lyubimov}

In general the CG representation has a number of configurational states lower than in the atomistic representation. As a results, the partition function of the system is modified. Even when the structure of the system is correctly represented by the pair correlation function in both the atomistic and the coarse-grained representations, some thermodynamic properties are different.  Because the number of configurational states that the system samples in the CG representation is reduced, the entropy of the CG system as measured in the simulation is also reduced with respect to the one in the related atomistic simulation. The free energy, on the other hand, is identical in the two representations, leading to the conclusion that  the internal energy also has to change during the process of coarse-graining. In order to develop a precise understanding of how these properties are modified during coarse-graining, it  is essential to build models that are based on a solid statistical mechanics foundation and analyze coarse-graining procedures by means of the tools of statistical mechanics.  

\section{Bottom-up and top-down models}
The coarse-grained models that have been proposed so far, mostly divide in two groups. The \textit{bottom-up} approaches start from the atomistic description and group atoms into new coarse-grained units. This procedure, if rigorously done, should provide information on the effective interaction potential between CG units. The potentials are  used as an input to the Molecular Dynamics simulations of the system represented by the CG description. 

In the bottom-up models the coarse-grained description is in most cases derived by matching specific physical quantities between the two levels of descriptions, atomistic and coarse-grained. Motivated by Henderson's theorem, which states that an isotropic potential which reproduces the correct pair structure of a fluid is unique up to a constant,  the physical quantity that is most commonly matched is the pair distribution function.\cite{Henderson}  This is the basis of the Iterative Boltzmann Inversion procedure.\cite{vandervegt,Muller-Plathe} Statistical mechanics directly relates the pair distribution function to the thermodynamic properties,\cite{HansenMcDonald} so that their proper agreement should be also ensured. Other approaches reproduce the data from atomistic simulations by matching the internal energy,\cite{JOHNSON} the forces,\cite{Voth} or simply minimizing the relative entropy.\cite{Shell}

The second type of approach is \textit{top-down} in the sense that the model is built to reproduce specific global properties of the system on the large scale (mesoscale) that characterizes the coarse-grained description. These methods are needed when there is not atomistic simulation that can provide a valid reference description for the numerical optimization. This happens when i) the system to be simulated is quite complex, for example in simulations of large biological objects, ii) or when the properties that are of interest are on a scale too large to be reached in atomistic simulations, for example dynamical properties close to a second order phase transition where the concentration fluctuation lengthscale diverges, iii) or when the atomistic description is itself imprecise, as can be the case in atomistic simulations of RNA and DNA, where, in some thermodynamic conditions, the potential is known to produce overstacking of the base-pairs and incorrect melting temperature.\cite{RNA,garcia,papoian} Belonging to the category of top-down approaches are all the mean-field theories of polymer structure and dynamics,\cite{polymer,doiedw} and models of membranes.\cite{brown}

 Necessarily top-down models are less precise than the bottom-up approaches for which a rigorous statistical mechanics procedure can guide the coarse-graining of the atomistic description. Top-down approaches are, however, quite reliable in reproducing the properties that are used in the construction of the model,\cite{pabloDNA} but are less reliable as far as other physical properties are concerned. For systems that are very complex, such as biological macromolecules and their complexes, for which the coarse-grained units cannot be straightforwardly defined, top-down models are the only feasible way to approach coarse-graining.

For some biological macromolecules such as proteins a number of bottom up approaches have been designed, mostly to study the dynamic of fluctuations around minima of energy and folding,\cite{Clementi,Wolynes,CGreview} while for nucleic acids the approaches have been mostly top-down because the atomistic forcefields need to be further developed.\cite{cheatham,cheatham1,garcia}

Our approach to the structure and dynamics of polymer liquids belongs to the group of the bottom-up methods, where the pair distribution function is reproduced across variable levels of coarse-graining, but differently from the methods discussed above the coarse-graining approach is largely analytical, and not numerical, and does not require the numerical fitting to atomistic simulations. 
	We also proposed a fine-grained model for the dynamics of proteins in solution based on a Langevin Equation for a chain of amino-acids represented as effective spheres. This model, which can be parameterized using either atomistic simulations or experiments, shows an excellent agreement with the time-correlation-functions measured in NMR relaxation experiments.\cite{Copperman}

\section{Modeling synthetic polymers as chains of coarse-grained units}
In recent years we have proposed a series of models to coarse grain the structure of macromolecular liquids.\cite{SAMB,Anthony,AnthonyPRL,YAPRL, YA2005,SAM2006} Our models concerns liquids of polymers that are isotropic and composed of $n$ molecules in a volume $V$, with each chain including a total number of monomer $N$. The density of chains is $\rho_{ch}=n/V$ and is related to the liquid monomer density $\rho=\rho_{ch}N$.
Every chain in the liquid is partitioned in a variable number of coarse-grained units or blobs, $n_b$, each containing a number $N_b=N/n_b$ of monomers, with the blob density $\rho_{b}=\rho/N_b$.

Starting from the Ornstein-Zernike integral equation we calculated the pair distribution function of the coarse-grained model and the effective potential acting between units. The approach is analytical. Using the potential we performed simulations of the coarse-grained systems, and  then compared thermodynamic quantities and structural quantities of interest from these coarse-grained simulations with united atom simulation data. The agreement between CG and atomistic descriptions is quantitative, while the direct correlation contribution at large distances, $c(k \rightarrow 0)=c_0$, is the only non-trivial parameter, which is evaluated either from experiments or from theory. Notice that atomistic simulations are not needed to parameterize the CG model, but they are only used to test the CG description.

Having an analytical solution brings some advantages to the method. The theory produces formal solutions of the static and dynamic quantities of interest and a formal analysis of how different properties are modified during the process of coarse-graining becomes feasible. Furthermore the approach is useful in multi-scale simulations of dense polymer systems with specific chemical structure because is system specific and reproduces the correct equation of state across various levels of coarse-graining.

The model applies to any type of polymer, because the lengthscale of coarse-graining is assumed to be larger than its local persistence length. By selecting a lengthscale larger than the persistence length, which is specific of the polymer considered, the coarse-grained units are statistically uncorrelated and follow a random walk in space. The chain of blobs can then be modeled as freely jointed, which affords an analytical formalism for the blob chain structure and interacting potentials. 
This model has unique characteristics because being analytical is fully transferable: it applies to different points in the phase diagram, and represents well any type of homopolymer liquid, independent of the molecular structure of the monomer.

In our method atomic-scale simulations are used only as a test and not to provide information to the coarse-graining approach. The purpose of developing coarse-graining methods is to have a mesoscale description that can be directly used in molecular dynamics simulations without the need of performing atomistic simulations to parameterize them. If atomistic simulations need to be performed at each thermodynamic condition of interest, there is no need of performing mesoscale simulations given that all the needed information would be already contained in the atomistic simulations. 

\begin{center}
\begin{figure}
\includegraphics[scale=0.5]{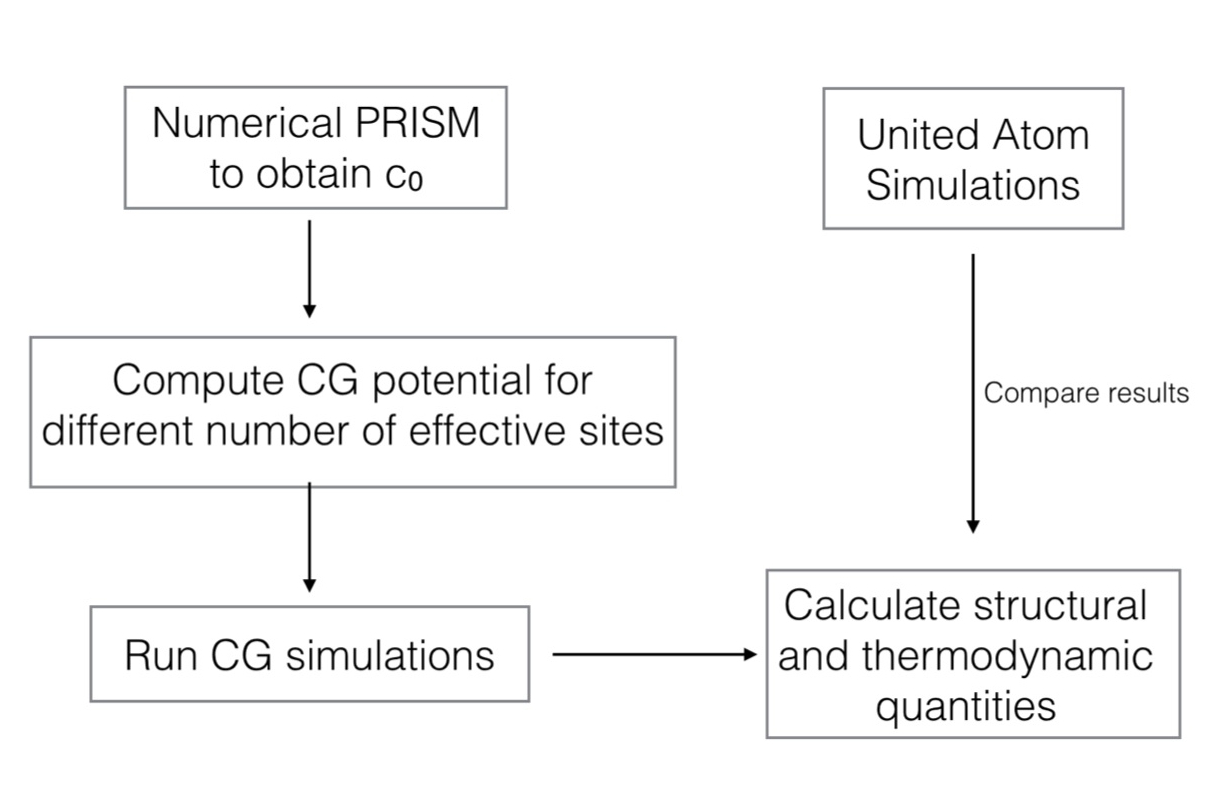}
\caption{Schematic outline of Integral Equation Theory Coarse-Graining Approach presented here. There are two different types of simulations: united atom and coarse-grained. The coarse-grained potentials are obtained from the analytical theory, and used in coarse-grained simulations with variable numbers of effective sites. Structural and thermodynamic quantities are then compared directly to united atom simulations. Reprinted with permission from Ref.\cite{Jaypaper}, Copyright [2014], AIP Publishing LLC.}
\label{uno}
\end{figure}
\end{center}

Figure \ref{uno} shows a schematic outline of our approach. We perform two
different types of simulations: united atom and coarse-grained. The coarse-grained
potentials, obtained analytically by solving the Ornstein-Zernike equation, are used in coarse-grained simulations with variable numbers of
effective sites. Structural and thermodynamic quantities are then compared
directly to united atom simulations. Tests have been done for polyethylene melts at a
variety of coarse-graining levels, and for systems in different
thermodynamic conditions and variable chain length.

The model has been extended to treat polymeric liquids where phase separation occurs, such as mixtures of polymers with different monomeric units and local semiflexibility, but also mixtures of polymers with the same molecular form but different degree of polymerization, and also melts of diblock copolymers with variable composition of the blocks.
Because the model has been solved analytically it is possible ot perform a precise analysis of the advantages and the  challenges of using a CG model. Some considerations emerge from this study that are more general than the model presented and relate to all CG models not just ours. In this way the discussion has a value beyond the simulation of liquids of polymers.

\subsection{Fine-graining approaches}
While the model we have studied and developed is of the type where the degree of coarse-graining is larger than the polymer persistence length, and belongs to the family of ultra coarse-graining models or molecular scale models, other coarse-graining models target a finer scale. These are the models that group together a small number of atoms to create a unit, for example they replace an aromatic ring with a bead\cite{Kremer}, or they group together the atoms in the side chain of each amino acid,\cite{Shea} or the represent each base in DNA with a reduced number of beads.\cite{LouisDNA}

Among these models, one that is general enough to be fully transferable to different polyolefins is the united atom model, which assembles together hydrogens and the related carbon atom in one bead. Units as $CH_3$, $CH_2$ and $CH$ all have different optimized values of the parameters that define their potential. The success of the United Atom (UA) model is due in large part to the small error that one can make when adjusting the parameters of the potential.\cite{UAmodel} As we show later on in this review, the error is small because the level of coarse-graining is minimal. The numerical optimization of the potential against atomistic simulations is feasible with negligible negative consequences on the precision of the calculations. Other models, which rely on a more extended level of coarse-graining, require to represent with a quite reduced number of parameters often a complex configurational space, with multiple free-energy barriers and pathways. 

Because this complex energy landscape can be influenced by the surrounding atoms, which can exert forces on the hidden atoms of the coarse-grained unit, the parameterization of the potential is hardly uniquely defined, but it depends on the position of the coarse-grained unit inside the molecule. In a nutshell the coarse-grained unit is not statistically independent of its surrounding atoms and it can be hardly treated as an independent statistical unit.
This is the basis of the problems that most coarse-graining approaches have in relation to the transferability and representablity of the potential.

\section{The problem of representability and transferability of coarse-graining models}
It is often noticed how coarse-graining models that at given state conditions of temperature and density, for example, are optimized to reproduce correctly a physical quantity such as the pair distribution function, cannot reproduce correctly other quantities, such as the pressure or the free energy of the system. In most cases CG models would need to be optimized for each given set of thermodynamic parameters, and for every type of molecular system that one wants to study. 

The lack of thermodynamic consistency between various representations is in fact a limitation that has delayed the widespread use of coarse-grained models in engineering and material science. For many numerical coarse-graining schemes, there is no guarantee that the resulting behavior of the coarse system will be consistent with what would have been observed by using a more detailed, more-expensive, model.\cite{JOHNSON,LOUIS,FALLER2008}

There is a second and more subtle reason that hampers the application of coarse-graining models. Numerically optimized CG models are built to reproduce faithfully the properties of the atomistic model against which the CG parameters where optimized. Even when considering all the possible physical properties of interest, and when the model is capable of reproducing those with very small errors, the quality of the coarse-graining model is limited by the quality of the atomistic simulation and by the possible lack of ergodicity of the system simulated. Because it is not know if the atomistic simulation is able to sample efficiently the free energy landscape, the quality of the coarse graining model is limited by the quality of the atomistic simulation.\cite{noe,methad,methad1} 

Motivated by these considerations, some recent studies have focused on understanding how well a less-than-optimal coarse-grained description, i.e. a CG model that contains errors, is still able to provide some useful guidance to the related atomistic simulations. In these studies CG models are used for sampling the configurational space and uncover deep energy minima in the atomistic potential, which would be otherwise not accessible to the atomistic simulation.\cite{Diner} The coarse-grained description can guide the system to sample a larger region in the configurational space than the one accessible to the atomistic simulation in the same time window.

Simulations that are run on the mesoscale are periodically stopped to rebuild the atomistic structure around the coarse-grained one, and then the simulation is carried on with atomistic resolution. Mesoscale and atomic scale simulations are alternated for a full and consistent sampling of the free energy landscape. The quality of the sampling depends on the reliability of the coarse-graining procedure and on the quality of the atomistic description. 

\section{Integral Equation Theory of Coarse-Graining}
In constructing a coarse-grained model it is common to start from an atomistic representation and to build the new coarse-grained description in a reduced representation, by matching the quantitative value of a physical property in the two descritpions, atomistic and reduced. Different theories match different quantities, specifically the most common quantities are the matching of the structural distribution, i.e. the pair distribution function, using the Inverse Boltzmann Iterative procedure\cite{IBI,IBI1}, the force matching procedure,\cite{Voth} to map the derivative of the mean-force potential, the mapping of the internal energy,\cite{vandervegt} and finally the search for a minimized information entropy.\cite{Shell} These methods are successful in reproducing quantitatively the physical properties that are initially fitted, but in most cases  they cannot ensure representability and transferability of the potential so derived.

We approached the problem from a different perspective. It is known that the pair distribution function is uniquely defined\cite{Henderson} and that the real pair distribution function, when is known, allows for the calculation of all the thermodynamic quantities of interest.\cite{McQuarrie} We derived the pair distribution function from the solution of integral equation theory for the coarse-grained description, starting from the integral equation theory in the atomistic representation, which is the PRISM approach.\cite{PRISM1} 

The solution of the pair distribution function from the Integral Equation theory does not require performing atomistic simulations.  Instead the atomistic pair distribution function is defined starting from the molecular and the thermodynamic parameters of the system. Once the pair distribution function is derived, then the effective potential between the coarse-grained units is calculated using the appropriate closure.

The derived potential is an input to mesoscale simulations of the coarse-grained system, which once they are performed provide all the molecular and thermodynamic quantities of interest. Those are, for example, the pair distribution function of the coarse-grained description, the pressure, excess free energy, as well as internal energy and entropy. The structural and thermodynamic quantities are then compared with the ones measured in atomistic simulations for the system under study. Atomistic simulations are used only as a test of the correctness of the proposed coarse-grained potential, and not to optimize in any way the coarse-grained description.


\section{Derivation of the effective coarse-grained potential in a liquid of polymers represented as chains of multiblobs}
In our multiblob coarse-grained model each polymer is described as a chain of $n_b$ soft blobs.\cite{SAMB, Anthony2, Anthony,AnthonyPRL} The number of soft spheres can be varied, starting from the single soft sphere representation,\cite{YAPRL,YA2005,SAM2006} going to the multiblob description. The only constraint is that the size of the blob has to be larger than the persistence length of the polymer, for polyethylene each blob needs to have at least $30$ monomers. This allows us to adopt a freely jointed chain model, as the chain of blobs will follow Markov statistics. By adopting this coarse-grained description the simulation can include  many more polymer chains than would otherwise be possible to simulate. 

Given $N$ monomers with a chain density $\rho_{ch}$, and an effective segment length $\sigma=\sqrt{6/N}R_g$ with $R_g$ being the polymer radius of gyration and $R_{gb}=Rg/\sqrt{n_b}$ the blob radius of gyration. The number of underlying monomers per blob is given as $N_b=N/n_b$. Coarse-grained or fictitious interacting sites are taken to be located at the center of mass of the polymer chain for the soft sphere model or at the centers of mass of several monomers along the same chain for the connected blob model. The relation between center of mass fictitious sites and real monomer sites is derived by solving a generalized matrix Ornstein Zernike equation.\cite{Anthony}

Single soft sphere representation, three, and five blob representations, have been formally derived. For the three-blob and five-blob representations the blobs are not all equivalent, as the blobs at the end of the chains are different than the one(s) in the internal part of the chain.  For chains with a large number of blobs, more specifically when there are more than five coarse grained sites per chain, end effects become negligible and it is possible to use a blob-averaged description.\cite{Anthony2}. 

The intramolecular distributions in the blob averaged limit are normalized as $\Omega(k)=\omega(k)/N$, for the blob-blob (bb), blob-monomer (bm), and monomer-monomer (mm) distributions.  The normalized blob-monomer and the blob-blob distributions are given by
\begin{equation}
\hat{\Omega}^{bm}(k) = \frac{1}{n_b} \left[ \frac{\sqrt{\pi}}{k R_{gb}}Erf\left( \frac{kR_{gb}}{2} \right)e^{-\frac{k^2R_{gb}^2}{12}} - 2 \left( \frac{e^{-n_b k^2R_{gb}^2}-n_b e^{-k^2R_{gb}^2}+n_b-1}{k^2 R_{gb}^2 n_b (e^{-k^2 R_{gb}^2}-1)} \right) e^{-k^2R_{gb}^2/3} \right] \ ,
\label{EQ:omegabavbmch5}
\end{equation}
and 
\begin{equation}
 \hat{\Omega}^{bb}(k) = \frac{1}{n_b} + 2\left[\frac{e^{-n_b k^2R_{gb}^2}-n_b e^{-k^2R_{gb}^2}+(n_b-1)}{n_b^2(e^{-k^2R_{gb}^2}-1)^2} \right] e^{-2k^2R_{gb}^2/3} \ .
 \label{EQ:Ombbkavch5}
\end{equation}
The monomer distribution $\Omega^{mm}(k)$ is normalized as 
\begin{equation}
\Omega^{mm}(k)=\hat{\omega}^{mm}(k)/N=\frac{2}{n_b^2k^4R_{gb}^4}(k^2R_{gb}^2n_b-1+e^{-n_bk^2R_{gb}^2}) \ .
\label{EQ:wmmkavN}
\end{equation}

Given the Ornstein-Zernike relation for the coarse-grained blob representation 
\begin{equation}
 \hat{h}^{bb}(k)= n_b \hat{\Omega}^{bb}(k) \hat{c}^{bb}(k) \left[ n_b \hat{\Omega}^{bb}(k) +\rho_b \hat{h}^{bb}(k) \right]\ ,
 \label{EQ:hbbkavch5}
\end{equation}
the direct correlation function is given as
\begin{equation}
\hat{c}^{bb}(k)=\frac{\hat{h}^{bb}(k)}{n_b \hat{\Omega}^{bb}(k)\left[n_b \hat{\Omega}^{bb}(k)+\rho_b \hat{h}^{bb}(k)\right]} \ .
\label{EQ:cbbkavch5}
\end{equation}

From the direct correlation function the interaction potential is calculated by evaluating the Hyper Netted Chain potential from the Fourier transform of Eq. \ref{EQ:hbbkavch5} and Eq. \ref{EQ:cbbkavch5} as
\begin{equation} 
\frac{v^{bb}(r )}{k_BT}=-\ln{[h^{bb}(r )+1]}+h^{bb}(r )-c^{bb}(r ) \ .
\label{EQ:Closure3}
\end{equation}
The solution can be performed either analytically or numerically.

When $|h^{bb}(r)|<<1$, which always holds at large separations ($r>>1$ in units of $R_{gb}$) and at any separation for representations with large $N_b$ and high densities, the potential further simplifies to
\begin{equation}
 V^{bb}(r) \approx -k_b Tc^{bb}(r) \ .
\end{equation}
This formula is referred to as the mean spherical approximation (MSA) in the literature, and applies to low compressible polymer liquids. If this formalism is improperly used to treat low density liquids, where the mean spherical approximation does not hold, this approximation would lead to unphysical behavior.

We now focus on the effective direct correlation function and the MSA potential for large separations in real space. In this limit, $r>> 1$ (in $R_{gb}$ units), the inverse transform integral is sufficiently dominated by $\hat{c}^{cc}(k)$ (in the small wave vector limit) that the large wave vector contribution can be entirely neglected.  Furthermore, since the expansion for small wave vectors is bounded at large $k$, the error incurred in using the small $k$ form with the integral bounds extended to infinity is small. This approximation leads,  for $r>>1$, to
 \begin{equation}
\begin{array}{rcl}
  c^{bb}(r) & \approx & \frac{-N_b\Gamma_b}{2 \pi^2 \rho_m R_{gb}^3 r} \int_0^{\infty} \left(k \sin(k r) \left[ \frac{45}{45+\Gamma_b k^4} +\frac{5k^2}{28} \frac{13 \Gamma_b k^4 - 3780}{(\Gamma_b k^4+45)^2} \right] \right) dk \\
   &=& \big[- \left(\frac{45 \sqrt{2} N_b \Gamma_b^{1/4}}{8 \pi \sqrt{3} \sqrt[4]{5} \rho_m R_{gb}^3}\right)\frac{\sin(Q^{\prime}r)}{Q^{\prime}r}e^{-Q^{\prime}r} +\left( \frac{\sqrt{5}N_b}{672\pi\rho_m \Gamma_b^{1/4} R_{gb}^3} \right) \big[(13Q^{3}(Q^{\prime}r-4))cos(Q^{\prime}r)
\\ &&+\left(\frac{945+13Q^{4}}{\Gamma_b^{1/4}}\right)r\sin(Q^{\prime}r)+\frac{945r}{\Gamma_b^{1/4}} cos(Q^{\prime}r)\big] \frac{e^{-Q^{\prime}r}}{Q^{\prime}r} \big] \ ,
\label{force-mb-ss-examples-b}
\end{array}
\end{equation}
where $Q^{\prime}=5^{1/4}\sqrt{3/2} \Gamma_b^{-1/4}$ and $Q \equiv Q^{\prime} \Gamma_b^{1/4}$ .
The key quantity of interest is $\Gamma_b=N_b \rho |c_0|$, which is defined once the molecular and thermodynamic parameters are known, but also depends on the direct correlation function at $k =0$, which is unknown.

The range of the potential, in units of the radius of gyration of the blob, scales as $N_b^{1/4}$. This scaling behavior describes how the interaction between effective units propagates through the atomistic sites in the macromolecular liquid. This pathway follows a random walk in the space defined by the lengthscale of the blob-blob interpenetration, which also scales with the degree of polymerization as $N_b^{1/2}$.

Interestingly, the range of the potential increases with the number of monomers comprised in the coarse-grained unit, i.e. blob or soft sphere, while the potential at contact decreases. However the interblob potential does not vanish even when the length of the polymer chain becomes infinity, indicating that intermolecular interactions between polymers are important even for infinitely long chains. This result disagrees with the conventional assumption in polymer physics that intramolecular interactions are dominant over intermolecular contributions, and a polymer melt can simply be described by mean-field approaches of the single chain.\cite{doiedw}

The potential becomes longer-ranged with increasing the lengthscale of coarse-graining, maximizing the gain in computational time. However the presence of long-ranged forces in the molecular dynamic simulations makes the use of a large box necessary, as the simulation box is usually chosen to be at least twice the range of the effective potential. The inconvenience of having long-range interactions can be alleviated by the use of simulations in reciprocal space, as it is conveniently done in the case of electrostatic interactions using the Ewald summation.\cite{HansenMcDonald}

The potential has long-ranged, slowly decaying repulsive component and a second attractive part that is smaller in absolute value than the repulsive part. This attractive contribution is important when one evaluates the thermodynamic properties of the system and cannot be discarded. Higher order terms, which are present in the equation of the potential, tend to give increasingly more negligible contributions: the potential in our simulations is often truncated after the first attractive well, or more rarely the second repulsive contribution, depending on error minimization.

It is interesting to note that the attractive contribution is present in the effective potential even when the intermolecular atomistic potential, from which the coarse-grained potential is derived, is purely repulsive. This indicates that the attractive contribution to the intermolecular potential is, at least partially, a consequence of coarse-graining and the propagation of the interactions through the liquid. Being the resultant of the projection of many-body interactions onto the pair of coarse-grained units, the attractive component of the potential is, at least in part, entropic in nature.

When the atomistic-scale interaction includes already an attractive part, i.e. for example the monomer-monomer interaction is a Lennard-Jones potential or a FENE potential, the latter provides a contribution to the total attractive component of the CG potential. In that case the attractive CG component is enhanced with respect to one arising from a "pure" hard-sphere monomer-monomer interaction. 

The CG potential becomes longer ranged with increasing level of coarse-graining, so that for models of the fine-graining type, the potential is still short-ranged and the error that can possibly occur in the thermodynamics when there are imperfections in the calculations of the pair distribution functions, $g(r)$, and the related potential, $v(r)$, is still small. It is for these models that the Iterative Boltzmann Inversion procedure becomes a promising strategy to calculate the interacting potential, which is optimized to reproduce the structure but also can predict the thermodynamics with small error. Other methods as well, which optimize the potential by optimizing the forces or by minimizing the information entropy, should be most precise when the number of atoms that are grouped together into the effective coarse-grained unit is small. 

In general, fine grained models are complex because they are very specific of the local structure and geometry of the unit that is being coarse-grained. Dihedral angles, branching, local angles are in most cases different in different monomeric units, and the potential derived from the optimization against atomistic level simulations is hardly applicable to similar units belonging to polymers that have different chemical composition.\cite{Peter} A typical example of this problem are coarse-graining models for proteins, where, even if the number of building blocks is reduced to only twenty amino acids, not only the position of each aminoacid inside the primary structure of the protein but also the chemical nature of its near-neighbor and next-near neighbor amino acids is important to correctly predict the properties of the protein, for example its folding. In this way a simple potential that is specific of pairs of aminoacid is limited in predicting quantitative and experimentally-consistent physical quantities, when they are related to a precise evaluation of the energy of the system.

\section {Thermodynamics}
From the pair distribution function the equation of state and the related thermodynamic quantities of interest are derived for our multiblob coarse-grained description. The normalized pressure is given by the virial expansion 
\begin{eqnarray}
\frac{P}{\rho_{ch}k_BT}= 1- \frac{N\rho c_0}{2} \ ,
\label{pzero}
\end{eqnarray}
with $\rho_{ch}$ the number chain density and $c_0 < 0$ the $k=0$ limit of the total correlation function. The isothermal compressibility is 
\begin{eqnarray}
\rho k_B T \kappa_T=\frac{N}{1-\rho N c_0} \ .
\end{eqnarray}
The excess Helmholtz free energy per monomer, calculated relatively to the energy of the system in its gas phase, is
\begin{eqnarray}
\frac{F-F_0}{nk_BT}=-\frac{Nc_0\rho}{2} \  .
\end{eqnarray}
These equations are derived from the approximated analytical solution of the effective potential, which is accurate in the mean-field limit of a nearly incompressible liquid.\cite{Likos} This equation of state holds for any level of coarse-graining in the multiblob description, while all the non-ideal contributions that arise from system-specific interactions are contained in the non-trivial parameter $c_0$. When deriving thermodynamic properties from the equation of state it is necessary to account for the state dependence of the parameter $c_0$ for the system under study. 

The \textit{internal} energy per chain, defined as the potential energy plus the kinetic energy, $U=E+K$, is found to have a more complex behavior. In the soft sphere representation, where $n_b=1$ and $N_b=N$,  the internal energy per chain is
\begin{eqnarray}
\frac{U}{nk_BT}=\frac{3}{2} -\frac{\rho N c_0}{2} \ ,
\end{eqnarray}
where the first term in the right-hand-side of the equation is the kinetic energy of the $n$ classical point particles, whereas the second term is the ensemble average of the potential energy arising from the intermolecular interaction contribution, which is identical in the soft sphere representation to the excess free energy. The related entropy per chain for the liquid of soft particles is given by the simple identity $U=F + TS$ as 
\begin{eqnarray}
\frac{S}{nk_B}=\frac{3}{2} +F_0 \ .
\end{eqnarray}
In the limit that the whole macromolecule is represented as a single-point particle, the entropy is only translational, no intramolecular configurational entropy is present in each coarse-grained site. 

In the underlying atomistic system, however, the entropy and the internal energy have additional contributions from the chain configurations, which are not accounted for in the soft sphere model. In the same way, those thermodynamic quantities in the multiblob description contain information from the multiblob chain configurations. Both entropy and internal energy are expected to depend on the level of coarse-graining.

The \textit{potential} energy in the multiblob description is composed of an \textit{intra}molecular and an \textit{inter}molecular contribution. The intermolecular component is calculated by a simple generalization of the soft sphere procedure as
\begin{eqnarray}
\frac{E^{bb}_{inter}}{nk_BT}=\frac{2 \pi \rho_{b}}{k_BT} \int_0^{\infty}v^{bb}(r)g^{bb}(r)r^2 dr \approx  -\frac{\rho N c_0}{2}  \ .
\end{eqnarray}
The dependence on the coarse-grained model emerges instead from the intramolecular contributions to the potential energy, 
\begin{eqnarray}
\frac{E^{bb}_{intra}}{nk_BT}=\frac{3}{2}(n_b-1)+\frac{1}{2}(n_b-2) \ ,
\end{eqnarray}
where $n_b=N/N_b$ is the number of blobs.

At a given temperature, while the excess free energy is constant, the number of degrees of freedom and the related entropy depend on the extent of coarse-graining. In this way, also the internal energy and the potential energy change with the number of internal degrees of freedom. Furthermore, the entropy correlates with the lengthscale of coarse-graining; the structure defined at a lengthscale larger than the lengthscale of coarse-graining is conserved, while the structure defined at a smaller lengthscale is averaged out. 

As the internal energy changes with the level of coarse-graining so does the specific heat, defined as $C_V=(\partial U/ \partial T)_V$. The specific heat directly depends on the number of degrees of freedom that are available to the system to store energy. When a molecule is represented with two different levels of coarse-graining the number of degrees of freedom changes and so does $C_V$.

The emergence of a phase transition, however, is determined by the free energy and the discontinuity in one of its derivatives with respect to the related thermodynamic variable. The value of the free energy as a function of the thermodynamic parameters does not change when the level of coarse-graining is modified, so that the phase diagram predicted by our coarse-graining model is identical, independent of the level of coarse-graining that is selected. 

We find interesting to notice that all the quantities that relate to the global/macroscopic properties of the liquid, such as the free energy and the pressure, do not depend on the choice of the lengthscale of the coarse-grained unit, but they depend only on the  thermodynamic parameters that are defined at the monomer level, namely the number of monomers in a chain, $N$, the monomer liquid density, $\rho$, and the direct correlation function at the macroscopic scale, $c_0$. This is correct, as the ``bulk'' properties of a liquid should not depend on the level of details employed to describe the molecules if the coarse-grained description is consistent.

\section{An universal equation of state for a variable-level coarse-grained representation of polymer melts}
In our model it is possible to take advantage of the fact that the excess free energy, compressibility, and pressure do not depend on the number of coarse-grained units in which the molecule is partitioned. We derived an equation of state for the polymer liquid starting from the simplest,  most reduced, representation, where the whole molecule is described as a point particle interacting through an effective long-ranged potential. This is the so-called ``soft-sphere model'', where the interaction potential is given by the solution of the integral equation for the center-of-mass of the molecule.

As mentioned earlier on, the only non-trivial parameter in our theory is the thermodynamic parameter $c_0$, which is related through the Ornstein-Zernike equation to the compressibility of the system and to the equation of state. This parameter is also independent of the degree of coarse-graining and can be conveniently calculated in the soft-sphere representation. 

The resulting equation of state is of the form of a Carnahan-Starling expression but includes numerical prefactors that reflect the chain connectivity and the fact that the real potential is not of the simple hard-sphere form
\begin{eqnarray}
\frac{P}{\rho k_B T}=\frac{4(\eta_{eff}+c_1 \eta_{eff}^2 + c_2 \eta_{eff}^3)}{(1-\eta_{eff})^3} \ .
\label{pressure}
\label{eqnstate}
\end{eqnarray}
The pressure is given as a function of the soft sphere packing fraction 
\begin{eqnarray}
\eta_{eff}=\frac{\pi}{6} \rho d^3 \ ,
\end{eqnarray}
and three parameters: an effective soft sphere diameter, $d$ , and two other parameters, $c_1$ and $c_2$, which are specific of the polymer under study.

\begin{center}
\begin{figure}
\includegraphics[scale=0.5]{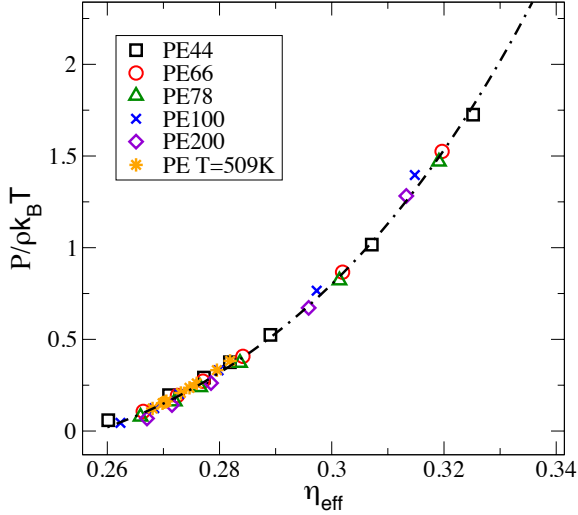}
\caption{Pressure as a function of the packing fraction for united atom simulations. Simulations carried out at constant temperature  $T=509 \ K$ and increasing polymer chain length are depicted with orange stars. All the other points are simulations at $T=400 \ K$ and variable densities and chain lengths. The dotted-dashed line is Eq. \ref{eqnstate}, which does not depend on the degree of polymerization, $N$. Reprinted with permission from Ref.\cite{Jaypaper}, Copyright [2014], AIP Publishing LLC.}
\label{due}
\end{figure}
\end{center}

Figure \ref{due} shows data for the normalized pressure as a function of the effective packing fraction for a number of united atom simulations of polyethylene performed at $T=400  \ K$ and increasing degree of polymerization ($N=44$, $66$, $78$, $100$, and $200$) and variable density,  as well as samples at $T=509 \ K$, density $\rho=0.03153 \ sites/A^3$, and  degree of polymerization  $N=36$, $44$, $66$, $78$, $100$, $192$, $224$, and $270$. All the samples fall onto an universal curve, which is well represented by the equation of state for soft spheres, Eq.\ref{eqnstate}.

In the same theoretical framework the direct correlation function at $k \rightarrow 0$ is expressed as a function of the parameters as
\begin{eqnarray}
c_0=-\frac{4\pi d^3}{3} \frac{1+c_1\eta_{eff} +c_2 \eta_{eff}^2}{(1-\eta_{eff})^3} \ .
\label{czero}
\end{eqnarray}

By plotting the normalized pressure as a function of the density for a number of atomistic simulations of polyethylene melts at varying temperature and degree of polymerization we see that the data follow an equation of state if plotted as a function of an effective packing fraction, $\eta_{eff}$. Once we assume that the parameters $c_1$ and $c_2$ are independent of $N$ and temperature, the optimized effective sphere diameter  is found to be $d=2.5 \ \AA$, while the other two  parameters are $c_1=-11.9$ and $c_2=31.11$. 

The simulations are found to reproduce quantitatively the trend of pressure as a function of density of the equation of state without any post-optimization scheme or fitting procedure, and across variable levels of coarse-graining. 

\subsection{Methods to evaluate the compressibility parameter $c_0$}
By using Eq. \ref{czero} we can estimate $c_0$ for any chain length at any temperature for polyethylene.
$c_0$  is the only parameter that does not describe in a trivial way physical or molecular quantities. Other parameters, besides the direct correlation $c_0$, are the thermodynamic properties of temperature, $T$, and density, $\rho$, as well as the structural properties of $N$, the number of monomers, and the effective segment size, $\sigma$. These parameters allow the method to be readily applied to a variety of polymers in variable experimental conditions.

A second possible procedure to calculate the monomer direct correlation parameter, $c_0$, is to solve the numerical solution of the PRISM equations with a realistic representation of the polymer chain. The solution of the PRISM equation provides results that are consistent with the equation of state method described above.\cite{Jaypaper}

An analytical equation for the $c_0$ parameter was obtained early on using the Gaussian thread model, which relies on the description of the polymer chain as infinitely long and infinitely thin, and is the model used to represent polymers in field theory.\cite{McCarty1}. While the PRISM thread model represents an idealized limiting case, it is not expected to give quantitative predictions for real chains of finite length and thickness. The analytical solution of the potential that has been discussed here does not rely on the use of the thread model.

A third method to evaluate the direct correlation function at large distances is to directly use the isothermal compressibility of the liquid under study, experimentally determined. The isothermal compressibility, $\kappa_T$,  which  is also preserved in coarse-graining, is related to the static structure factor $\hat{S}(k=0)$ as 
\begin{equation}
\rho k_B T \kappa_T= \hat{S}(k=0)=[N+\rho \hat{h}^{mm}(0)] \ ,
\label{EQ:compress}
\end{equation}
with $\hat{S}(k=0)$ the $k \rightarrow 0$ limit of $\hat{S}(k)=[\hat{\omega}^{mm}(K)+\rho \hat{h}^{mm}(K)]$. The isothermal compressibility in the blob description is identical to the isothermal compressibility in the monomer description.
The value of $c_0$ is then determined, for example in the monomer description, as 
\begin{equation}
c_0=\frac{\hat{h}^{mm}(0)}{\rho N\hat{h}^{mm}(0) +N^2} \ ,
\label{EQ:co}
\end{equation}
with $\hat{h}^{mm}(0)$ related to the isothermal compressibility through Eq. \ref{EQ:compress}.

\section{Testing the theoretical predictions of the coarse-grained model against atomistic simulations}
To test the quality of the theoretical predictions we performed a series of simulations of soft-blob coarse-grained models with variable level of coarse-graining and united atom models, under the same set of molecular and thermodynamic conditions, for a variety of chain lengths and densities at two different temperatures. The united atom simulations do not provide  information to the coarse-grained model, but they are used to test the consistency of the coarse-grained models. 
The calculations start from the equation of state from which all the thermodynamic properties of interest are derived and then compared to the results from simulations following the scheme of Figure \ref{uno}.

\begin{center}
\begin{figure}
\includegraphics[scale=0.5]{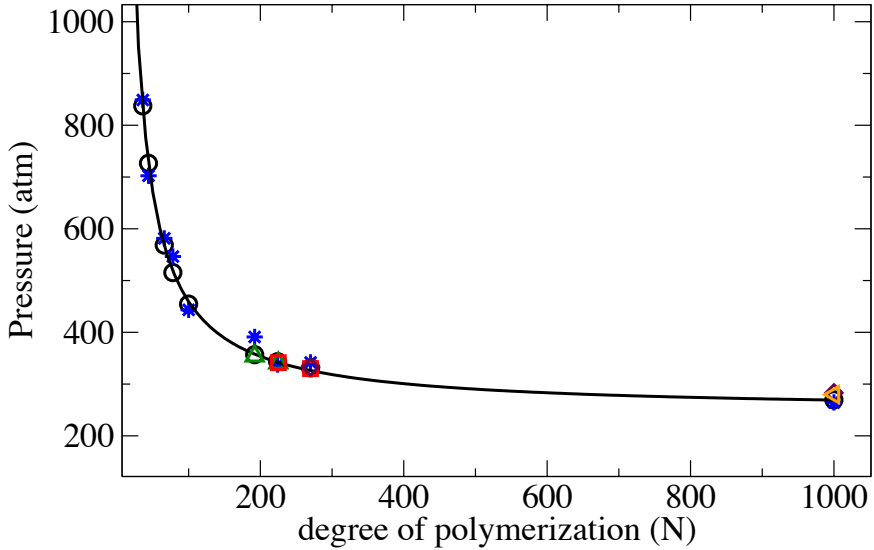}
\caption{Pressure as a function of the degree of polymerization for polymer liquids represented at different levels of coarse-graining. A hierarchy of soft-blob simulations are compared to atomistic simulations and to the analytical theory, Eq.\ref{pzero} with $c_0$ from Eq.\ref{czero}. The simulations were carried out at constat temperature, $T=509 \ K$, and constant density, $\rho=0.733 \ g/ml$. United atom simulations are represented by black circles, soft sphere model by blue asterisk, tri-blob by grees triangles, penta-blob by red squares, 10-blob by maroon diamonds, and 20-blob by orange left-oriented triangles. Reprinted with permission from Ref.\cite{Jaypaper}, Copyright [2014], AIP Publishing LLC.}
\label{tre}
\end{figure}
\end{center}	
	
As we expected, we see that macroscopic properties of the polymeric liquid do not depend on the lengthscale we select to coarse-grain the macromolecules, while the internal energy and entropy are model dependent.\cite{Jaypaper}  See for example in Figure \ref{tre} the calculation of the pressure as a function of the degree of polymerization for a set of simulations with variable degree of coarse-graining. The pressure shows consistency for all the samples; the data are also in agreement with the analytical expression of the equation of state.

\begin{figure}
\includegraphics[scale=0.5]{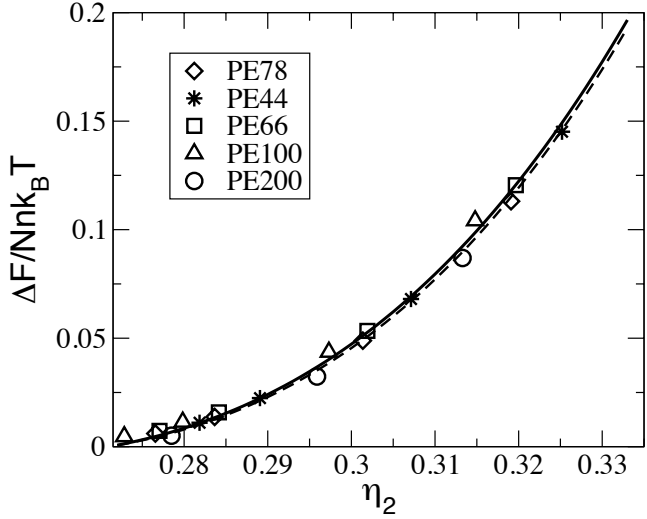} \\
\caption{Helmholtz free energy changes as a function of the packing fraction compared to a reference packing fraction of $\eta_1=0.27$, which is the lowest packing fraction that was simulated. All systems collapse to an universal curve, within numerical precision and independent of the degree of polymerzation. Solid and dashed lines are Eq.\ref{EQ:Helmholtz} for $N=44$ and $N=200$, respectively. Reprinted with permission from Ref.\cite{Jaypaper}, Copyright [2014], AIP Publishing LLC.}
\label{freee}
\end{figure}

The Helmholtz free energy per monomer is obtained through thermodynamic integration of the pressure as a function of the packing fraction
\begin{eqnarray}
\frac{F^{ex}}{Nnk_BT}  & = & \int_{\eta_1}^{\eta_2} \left(\frac{P}{\rho_{ch} k_B T}     \right) \ , \\ &=& \frac{-2(1-c_1-3c_2)\eta_{eff}^2+4 (1-c_2) \eta_{eff}}{(1-\eta_{eff})^2}- 4 c_2 \ln(1-\eta_{eff}) \ . \nonumber
\label{EQ:Helmholtz}
\end{eqnarray}
The excess Helmholtz free energy, associated to the liquid as distinct from the ideal gas, is shown to be independent of the degree of coarse-graining, as it is also illustrated by the example in Figure \ref{freee}.\cite{Jaypaper}

The excess Gibbs free energy in the canonical ensemble is calculated in the mean field approximation, which holds at liquid density, as 
\begin{eqnarray}
G^{ex}=\frac{n \rho}{2 N_b} \int v^{bb}(r) g^{bb}(r) d \textbf{r} \ ,
\end{eqnarray}
with $N_b$ the number of monomers per blob, from which the excess free energy per monomer 
\begin{eqnarray}
\frac{G^{ex}}{n N k_B T}= - \frac{\rho c_0}{2} \ .
\end{eqnarray}
It is worth noticing that the expressions for the Helmholtz and the Gibbs excess free energies are different because their calculations account for the different thermodynamic parameters that are controlled. The Helmholtz free energy is calculated at constant volume by integration over the packing fraction and so the density; these calculations account for the density dependence of the direct correlation function, $c_0$. The Gibbs free energy, instead, is calculated in the canonical ensemble, where the density is constant, and so is $c_0$.
For this quantity we evaluated the pressure dependence, input to 
\begin{eqnarray}
\frac{\Delta G}{nN k_B T}=\frac{1}{\rho k_B T} \int_{P_1}^{P_2} d P \ ,
\end{eqnarray}
by calculating the change in free energy related to the change in pressure observed when simulations at constant volume and temperature are performed as a function of the number of monomers per chain, and then compared to our analytical expression
\begin{eqnarray}
\frac{\Delta G}{nN k_B T} = [ \frac{1}{N}- \frac{\rho c_0}{2}]_2 - [ \frac{1}{N}- \frac{\rho c_0}{2}]_1 \ .
\end{eqnarray}
The excess free energy in both ensembles is a constant quantity and does not depend on the coarse-gained model adopted.

\subsection{Potential energy}
The internal energy, and the related potential energy, display a more subtle dependence on the number of coarse-graining units selected to partition the molecule.
For the \textit{soft sphere} description, for which $n_b=1$ and $N_b=N$, the potential energy includes only intermolecular contributions and is equivalent to the excess free energy 
\begin{equation}
\frac{E^{soft \ sphere}}{nk_BT}=\frac{2 \pi \rho_{chain}}{k_BT}\int_0^\infty v^{ss}(r)g^{ss}(r)r^2 dr\approx -\frac{\rho N c_0}{2} \ .
\label{EQ:Emb}
\end{equation}

\begin{figure}
\includegraphics[scale=0.5]{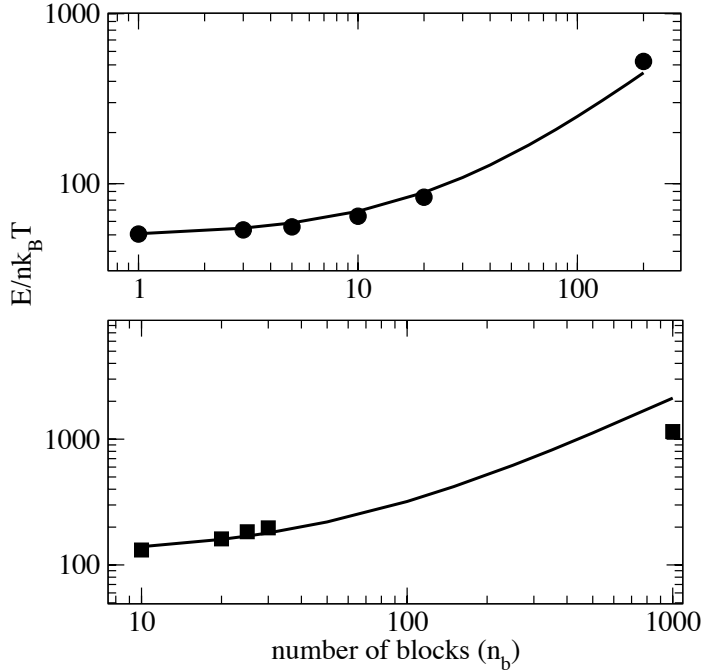} \\
\caption{Potential energy changes as a function of the number of blobs, for two samples. The firs is polyethylene with $N=200$ (top) and the second is polyethylene with $N=1000$ (bottom). The last point is the potential energy from united atom simulations. The solid line is the theoretical prediction, and is extrapolated to the last point in the limit of large $n_b$. Reprinted with permission from Ref.\cite{Jaypaper}, Copyright [2014], AIP Publishing LLC.}
\label{eee}
\end{figure}

In the multiblob description the potential energy has both inter- and intra-molecular contributions. The intermolecular part is calculated as an extension of the formula for the soft sphere
\begin{equation}
\frac{E^{bb}}{nk_BT}=\frac{2 \pi \rho_{b}}{k_BT}\int_0^\infty v^{bb}(r)g^{bb}(r)r^2 dr\approx -\frac{\rho N c_0}{2} \ ,
\label{EQ:Emb1}
\end{equation}
and gives a contribution that even in the multiblob description is a constant. The potential energy, however, contains in this case also contributions from the intramolecular structure, such as bond stretching, angle bending, torsional rotation and non-bonded pair interactions, which are representation dependent. 
Since the bond energy is a harmonic potential with a Gaussian probability distribution, the average bond energy is simply the equipartition result, 
\begin{equation}
\left<\frac{E_{bond}}{n k_BT}\right>=\frac{3(n_b-1)}{2} \ .
\label{EQ:Ebond}
\end{equation}
For the angular contribution to the energy we add an additional $E_{angle}\approx(n_b-2)/2k_BT$ contribution per chain. The total predicted energy is shown in the Figure \ref{eee} and represented by the line. 

From the simulations the potential energy is calculated as the average total energy minus the kinetic energy contribution. In the figure we show the potential energy per molecule for two different systems, PE200 at $\rho=0.8 g/mL$ at $T=400K$ and PE1000 at $\rho=0.733 g/mL$ and $T=509K$ as a function of the number of effective sites, $n_b$. The united atom simulations are represented by the last data point on the right of the figure, where the number of effective sites is equal to the number of united atoms. 

The figure shows that in both sets of simulations, the potential energy changes as the number of sites is increased, and that the agreement between theoretical expressions and simulations is quantitative up to the atomistic description where a small error is observed. In this case, however, the small disagreement observed between theory and simulations is due to the approximation of using the mean-field equation, $g^{bb}(r)\rightarrow 1$, which becomes increasingly less accurate as the local structure becomes important.

\subsection{Entropy}
The basic procedure of any coarse-graining formalism is the averaging of the microscopic states that are then represented by effective units, with the consequence that  the entropy of the system in a given coarse grained representation is different with respect to the atomistic description. This is  a direct consequence of the fact the coarse-graining reduces the dimensionality of the configuration space and smoothens the probability distributions. The extent of the change in entropy depends on the level of detail maintained in the coarse-grained representation, which determines the number of atomistic configurations that can be mapped into a single coarse-grained configuration. It can be quite large when the level of coarse-graining is extreme and the underlying chain is flexible.  This is commonly called as the ``mapping entropy,'' and is simply the difference in entropy of the atomistic model when viewed from the atomistic configurations and the coarse-grained configurations.\cite{Noid}

If the chain is assumed to have a statistical distribution of monomers in space that follows a Gaussian form, the entropy associated with increasing the number of blobs in the coarse-graining procedure
is given  as
\begin{equation}
\frac{S_{bb}}{nk_B}\approx \frac{3}{2}n_b+\frac{3}{2}(n_b-1)-\frac{3}{2}(n_b-1) \ln{\left(\frac{3 n_b}{8\pi R_g^2}\right)}+\ln \left(\frac{Ve}{\Lambda^3 n}\right).
\label{EQ:Sbbapprox}
\end{equation}

The first two terms in Eq. \ref{EQ:Sbbapprox} arise from the kinetic energy and bond potential energy, while the final two terms are the ideal translational and vibrational free energy. Importantly, there is no contribution in Eq. \ref{EQ:Sbbapprox} from the potential or $c_0$, since the increasing entropy with the number of blobs $n_b$ is due solely to the increasing configurational degrees of freedom and not the interaction potential itself.

Another type of entropy of interest is the relative entropy.\cite{Shell} This function is based on the ``information'' that is lost during corse-graining, which has to be minimized to optimize the coarse-grained model. Our coarse-grained formalism, based on liquid state theory,  is devised to reproduce the correct distribution function, so that the relative entropy between the coarse-grained sites and monomer sites is minimized, and the potential is optimized, without need for any variational approach. This is equivalently to say that the relative entropy, whch is based on the information function that discriminates between coarse-grained configurations sampled in the two levels of representation, is zero.

\section{Insights in coarse-graining from the thermodynamically consistent coarse-grained model}
One of the advantages of having an analytical method of coarse-graining is that it is possible to study how physical inconsistencies can arise from a selected coarse-graining procedure. More precisely it is possible to see how errors in the procedure can lead to errors in the resulting simulated quantities on the mesoscale.\cite{Anthony2}

An issue often reported when performing simulations of coarse-grained systems is that the mesoscale simulation describes a liquid that is too compressible in comparison to the more realistic modeling of the related atomistic simulations. This is not a problem for our model, which is largely analytical, but it affects numerically optimized coarse-graining methods such as the Iterative Boltzmann Inversion (IBI) procedure.

The conventional IBI procedure optimizes the coarse-graining potential by minimizing the disagreement between the atomistic and the mesoscale pair distribution functions, $g(r)=h(r)-1$, or equivalently the total correlation function, $h(r)$. When the effective potential is calculated from $g(r)$, the Iterative Boltzmann Inversion procedure rapidly converges to a total correlation function indistinguishable from the one the procedure is started with. The pressure, however, resulting from the mesoscale simulation that uses the derived potential, is found to be reduced from the correct value of the atomistic simulation. 

\begin{figure}
\includegraphics[width=3.5in]{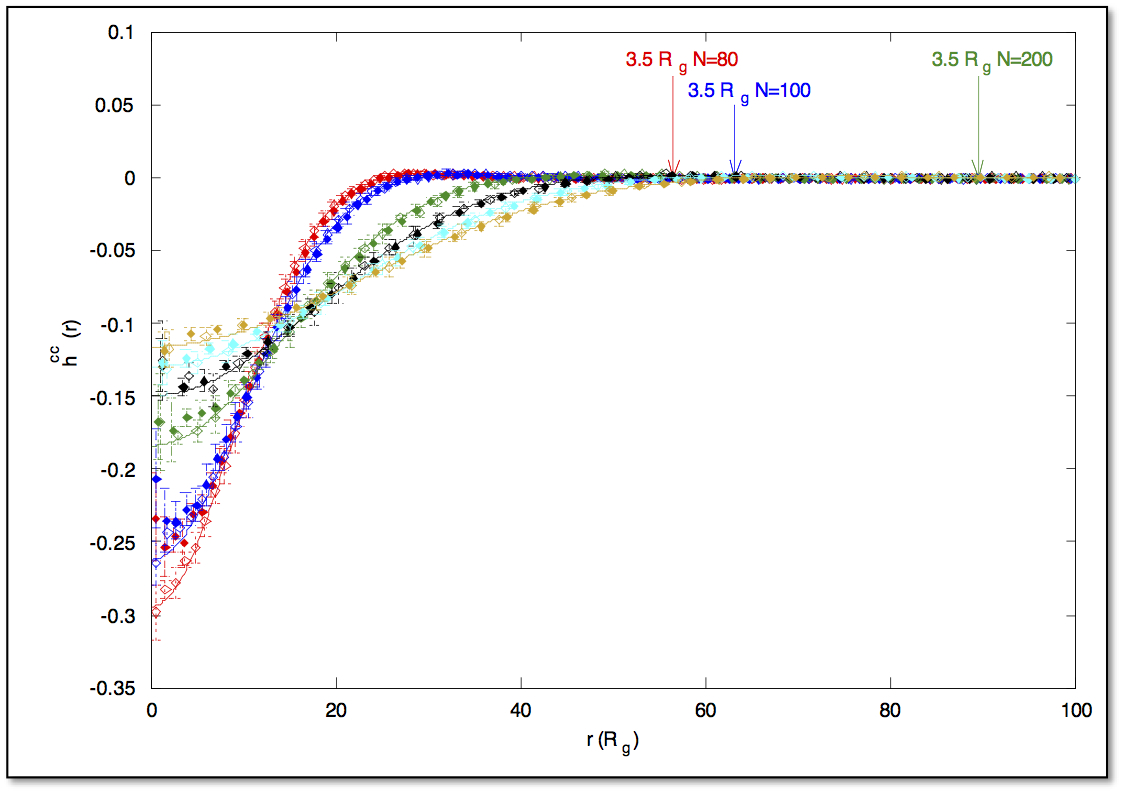} \\
\caption{Total correlation function for the soft sphere representation as a function of polymer center-of-mass distance, for chains of increasing degree of polymerization: $N=80$ is depicted in red, $N=100$ is blue, and $N=200$ is in green. The total correlation function from atomistic simulations (filled symbols) is compared to the theory (lines) and to the results from mesoscale simulations of the coarse-grained system, where the soft-sphere potential is derived by cutting $h^{cc}(r)$ at $r=3.5 \ R_g$. All three produce the same total correlation function.}
\label{hofrcut}
\end{figure}

To study the reason for the observed disagreement we started by comparing our analytical total correlation function for the soft sphere representation  to data from simulations. The agreement between analytical theory and atomistic simulations is quantitative (see Figure \ref{hofrcut}).

In the IBI the total correlation function against which the potential is optimized is defined up to a given interparticle distance, $r_{cut}$, which is a fraction of the box size in the atomistic simulation. To mimic the IBI we simply set the total correlation function to be identical to zero at a distance $r'$ that is larger than a given $r_{cut}$, where we select for the value of the cutoff distance $3.5 R_g$, with $R_g$ the radius of gyration of the polymer chain. Outside the radius of gyration of a polymer, for $r > R_g$, the probability of successfully find another polymer chain is $\approx 100 \%$ favorable and $h^{bb}(r) \approx 0$ is a valid approximation. Even more so at a distance as large as the one we selected.

\begin{figure}
\includegraphics[width=3.5in]{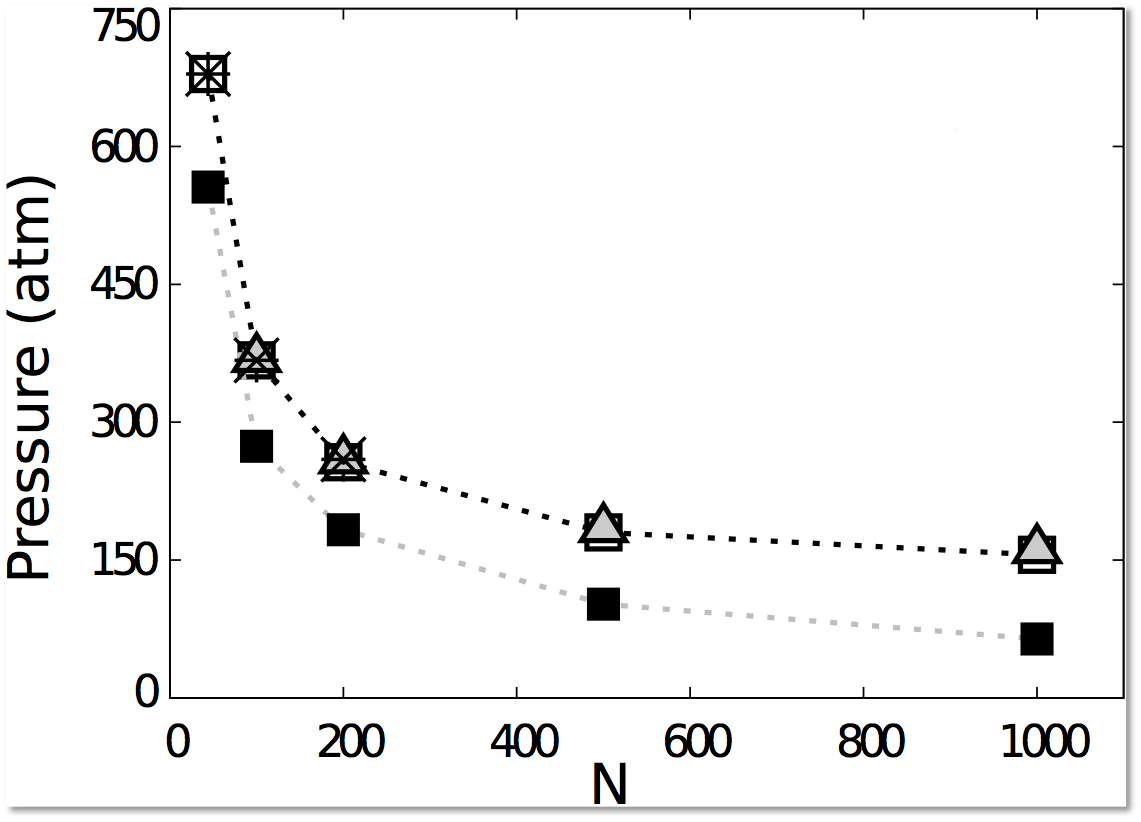} \\
\caption{Pressure measured from molecular dynamics simulations of the coarse-grained system performed either using the full tail of g(r) in the effective potentials (open squares) or the g(r) set to be equal to one for $r > 3.5 \ R_g$ (filled squares). Also shown are United Atom simulations (stars) for systems where they are available ($N \le 200$). Data are for systems consistent with Figure \ref{hofrcut}. Despite both potentials reproducing the structure with high accuracy, the pressure is clearly affected by cutting the pair distribution function at large distances. Reprinted with permission from Ref.\cite{Anthony2}, Copyright [2013], AIP Publishing LLC.}
\label{pressurecut}
\end{figure}

	Once the total correlation function is optimized with data up to $r=3.5 \ R_g$, we derive the potential and run molecular dynamic simulations for the soft-sphere liquid. The total correlation function obtained in the mesoscale simulation is indistinguishable within numerical error from the results of the atomistic simulations.
	However the pressure in the atomistic and mesoscale simulations is  different. Figure \ref{pressurecut} presents the calculation of the pressure with and without truncation of $h^{cc}(r)=g^{cc}(r)-1$. The error due to the truncation of the total correlation function at large distance leads to large errors in the pressure, because in the equation the pair correlation function is weighted by the distance elevated to the third power: small errors in the tail of a long-ranged potential strongly affects the precision of the equation of state. By truncating the total correlation function the potential that results from the procedure is also truncated and the pressure of the coarse-grained simulation is underestimated. As the equation of state is different when the pair distribution function is truncated, all the thermodynamic quantities that are derived from the equation of state will also be plagued by errors.

\begin{eqnarray}
\frac{P}{\rho_{ch}k_BT}=1-\frac{2 \pi \rho_{ch}}{3 k_B T} \int_0^{\infty} g(r) \frac{dv(r)}{dr}r^3 dr \ .
\label{pressure}
\end{eqnarray}

The error is larger the longer the range of the potential, the larger the lengthscale of coarse-graining, and the higher the density of the liquid. 
Unfortunately those are the conditions where coarse-graining is most useful. The range of the long repulsive tail of the potential increases with the level of coarse-graining and large-scale precision becomes important for polymer liquids when the chains are highly coarse-grained. Fine-gained models have smaller errors in the thermodynamics, but for these models the gain in computational time is limited.

Interestingly, even if the potential incorrectly predicts the thermodynamic properties the structure of the liquid in the form of the total correlation function is correctly reproduced. This is because the pair distribution function, and its related total correlation function, are notoriously insensitive to small differences in the potential. It is known that different potentials can lead to the same pair distribution function. For example, the total correlation function of a Lennard-Jones liquid is very similar to the total correlation function of a liquid of hard spheres of the appropriate diameter. In fact, pair distribution functions are mostly defined by the repulsive part of the potential, while the attractive contribution has little or no influence.

\section{Reconstruction of the dynamics: why the dynamics is too fast in coarse-grained models}
While structure and thermodynamic properties are well described across multiple levels of coarse-graining, the dynamical properties measured in mesoscale simulations of coarse-grained systems are always too fast when compared to the related atomistic simulations. For example the mean-square displacement and diffusion coefficient of a coarse-grained representation can be  several order of magnitude faster than in the atomistic  simulations. 


In the coarse-grained model local degrees of freedom are averaged out, and the molecules move rapidly over the simplified free energy landscape. While the system explores  the reduced configurational landscape, the measured dynamics is artificially sped up by the smoothness of the potential. The speed up in the dynamics is proportional to the level of coarse-graining, so that the largest computational gain is obtained for the most coarse-grained system.  

Given that the measured dynamics is unrealistically fast, it needs to be properly rescaled \textit{a posteriori} to recover the correct dynamics of the atomistic description. We have proposed an analytical procedure to rescale the mesoscale dynamics. Because all the structural and thermodynamic quantities are known in our model, it has been possible to derive from first principles a procedure for the reconstruction of the dynamics that is based on the solution of the Generalized Langevin Equations for the atomistic and the coarse-grained representations. We identified two steps in the rescaling procedure of the coarse-grained dynamics to reconstruct the atomistic description. A first rescaling aims at including the dissipation of energy due to the internal vibrational degrees of freedom into a time rescaling. Those degrees of freedom are averaged out in the soft-colloid representation during the coarse-graining process. The second step accounts for the change in shape, molecular surface exposed to solvent, and friction coefficient of the polymeric units as a consequence of coarse-graining.\cite{Lyubimov,IVAN1,IVAN2}

The procedure of dynamical reconstruction that we proposed is different from the usual one because it does not require performing atomistic simulations, and once the only parameter, i.e. the effective hard-sphere diameter of the monomer, is defined the theory is fully predictive. The  common strategy  is based, instead, on the application of a ``calibration curve" previously obtained through the numerical fitting of dynamical quantities: in the calibration curve procedure, parameters are optimized until one achieves the agreement for the dynamical properties of interest calculated in the atomistic and in the mesoscale simulations.\cite{Nielsen,Kremer}  However, the same reasoning that applies for static properties also applies in the case of the dynamics: once the atomistic simulations are performed it is not obvious the need of performing mesoscale simulations.

The numerical calculation of optimized calibration curves for the dynamics is quite difficult to achieve for macromolecular systems because the dynamics is mode dependent: there are in principle $N$ internal modes in any molecule formed by $N$ units and the degree of polymerization of a long chain can be of the order of one million monomers. Numerically optimized parameteric quantities are in general not transferable between systems in different thermodynamic conditions or with different chemical structure or increasing degree of polymerization. To overcome this problem, it is common to select coarse-grained units  that are very close in size to the atomistic units, so that the needed corrections to reach consistency in static, thermodynamic, and dynamic properties is minimal. In this case, corrections to the measured dynamics can be evaluated through a perturbative formalism, which should rapidly converge to the desired value. The resulting computational gain is, however, limited. 

Our procedure has been developed so far for the soft sphere representation, which affords the largest dynamicals gain. The  same principles hold for variable levels of multiblob representations. In the soft sphere represenation the internal dynamics cannot be studied, but the coarse-grained simulation can provide information on the center-of-mass diffusion.  The mesoscale simulations of the coarse-grained system provide data that once are properly rescaled have shown to predict center-of-mass dynamics in quantitative agreement with experiments and atomistic simulations. Because the soft-sphere representation is the one with the highest level of coarse-graining it requires the largest correction to the measured dynamics and is the one where possible errors in the procedure can have the most visible consequences.The fact that the agreement between the dynamics reconstructed from the coarse-grained simulations and the atomistic-scale representation is good suggests that our procedure is robust.

The ``entropic'' and ``frictional'' corrections to the mesoscale dynamics enter the Langevin equation of the coarse-grained system, and rescales its dynamics.  In the Langevin equation the energy dissipated is calculated by adopting a bead-spring monomer representation of the polyethylene chain.  The correction term that must be included in rescaling of the dynamics of coarse-grained simulations to account for the missing entropic degrees of freedom, starting from a freely-rotating-chain model, is equivalent to Equation \ref{EQ:Ebond}.\cite{IVAN1,IVAN2}

The second rescaling addresses the change in the polymer friction coefficient due to the reduction of surface  when the chain is coarse-grained. This rescaling factor is derived from the solution of the memory functions in the Langevin equations describing the dynamics of the polymer chains in the two levels of representation.

Using the proposed rescaling procedure, dynamical data from mesoscale simulations of polyethylene melts were compared with the ones measured in atomistic simulations and in experiments. The agreement between predicted and known properties was found to be almost  exact. Furthermore the procedure allowed for the prediction of new values of dynamical parameters, i.e. diffusion coefficients, for systems that were not yet simulated or measured experimentlally.

It is possible to take advantage of the artificial acceleration of the dynamics when coarse-grained representations are used to rapidly reach  an equilibrated state before starting an atomistic molecular dynamics simulation. In that case the variable level of coarse-graining allowed by our model is used to seamlessly change the resolution in coarse-graining. 

\section{A coarse-grained method for Protein Dynamics: the Langevin Equation for Protein Dynamics (LE4PD)}
We conclude presenting a coarse-grained approach to describe the fluctuating dynamics of folded proteins in dilute solutions. Many of the concepts presented in this chapter were used in building this coarse-grained model for the dynamics of proteins.\cite{Copperman}

Because we are interested in the motion of a large macromolecule immersed in a liquid of small water molecules and ions, the difference in size and in the characteristic timescale of the dynamics between solute and solvent suggests that it is appropriate to treat the protein and the solvent with different coarse-grained formalisms. The protein is simply treated as a collection of units centered on the position of the alpha-carbons and connected through effective springs along the primary sequence of the protein. The dynamics of each unit is coupled to the remaining others through effective pair potentials. The solvent is treated as a continuum, following the tradition of polymer physics. The solvent affects the dynamics of the protein through the viscosity, the random collision with the protein, and the friction.

In this model, the dynamics follows a modified Rouse-Zimm  Langevin Equation that we named Langevin Equation for Protein Dynamics (LE4PD). Inter-protein interactions are not important because the solution is diluted. Each coarse-grained macromolecular unit represents one amino acid, and the specific shape of the side-chain is accounted for through the hydrodynamic radius of the residue, its friction coefficient, and the effective interaction with others units. Because the level of coarse-graining is contained, and each unit represents a relatively small number of atoms, a numerical evaluation of the potential from atomistic simulations or experiments is appropriate: the resulting error is small and can affect only slightly the large-scale properties of the protein.  In the hydrophobic core the hydrodynamic interaction is screened.

In several points the LE4PD departs from the traditional model for polymer dynamics. The LE4PD approach uses a harmonically coupled description, but complete with site-specific dissipation, hydrodynamic coupling, and barriers to internal fluctuations, calculated directly from the structural ensemble created by the simulation of the protein in aqueous solvent. The knowledge of the roughness of the free energy landscape, i.e. the sampled energy barriers, provides information on the long-time dynamics. The predictions for the time correlation functions calculated from the theory exceed in timescale the time correlation function directly calculated from the simulation trajectories. Assuming that in a longer timescale the protein still samples the configurational space covered by the simulations, then the motion described by our approach in its present form ensures an accurate determination of the dynamics over a wider range of timescales than the simulation itself. 

The approach is based on the fundamental picture of proteins as heterogenous polymers which are collapsed into a definite tridimensional structure, which nevertheless retains some amount of flexibility. As opposed to a rigid body, where the global modes are the only degrees of freedom in the system, protein dynamics include both rotational and internal fluctuation modes. Our description accounts for internal dissipation due to fluctuations in the hydrophobic region by accounting for an effective protein internal viscosity and considering the relative exposure of each amino acid to the hydrophobic region. With the correct dissipation, the linear modes of harmonically coupled objects provide a simple but accurate description of the fluctuations of the molecule. The theory predicts local dynamics in close  agreement with experimentally measured time correlation functions, such as $T_1$, $T_2$, and $NOE$ data from NMR experiments.\cite{Copperman}

\section{A Summary of the Main Points in this Review}

In our approach the coarse-graining of macromolecular systems is based on the Integral Equation theory; pair distribution functions and effective potentials for the coarse-grained units have been calculated for models that have variable levels of resolution. The approach affords the analytical solution of the potential, from which we developed an analysis of the structural, thermodynamic, and dynamical properties of the coarse-grained description as a function of the resolution of the model, or level of coarse-graining. Structural and thermodynamc properties between the persistent length resolution to the most extreme level of coarse-graining, where the whole molecule is represented as a point particle interacting through a soft long-ranged potential, are properly described by our coarse-grained approach. 
For resolution smaller than the polymer persistence length, the general properties of our coarse-grained model still hold, but the solution of the integral equation theory has to be performed numerically. We expect that thermodynamic and structural consistency is maintained also on the very local scale.

Our theory differs from most alternative coarse-graining approches because it does not require performing high-resolution  simulations to numerically parameterize the coarse-grained model. Atomistic simulations are used only as a test of the consistency of the  coarse-grained description. We find that our model reproduces the correct pressure, structural distributions, compressibility, and free energy of the underlying system. Internal energy and entropy are instead depending on the degree of resolution of the coarse-grained model.
  
All the quantities depend on one non-trivial parameter, i.e. the total correlation function at $k \rightarrow 0$, $c_0$, which can be directly determined from the experimental isothermal compressibility of the liquid. This parameter is system specific, depends on the thermodynamic conditions, but does not depend on the resolution of the selected coarse-grained description. 

In numerical procedures of optimization it is common to rely on pair distribution functions obtained either experimentally or from atomistic simulations; in both cases the function is truncated at large distance. We have shown that the truncation leads to a consequent error in the thermodynamics of the coarse-grained simulation.

Coarse-graining also produces a speeding up of the dynamics that allows for the fast simulations of molecules within a reduced description. The dynamical properties, however, in the coarse-grained simulations are unrealistically accelerated and need to be properly rescaled to give realistic values of the dynamics. We have shown that both entropic rescaling of the degrees of freedom and rescaling of the effective friction coefficients are important to reconstruct the correct dynamics. Global dynamics and diffusion coefficients are well predicted with our rescaling procedure applied \textit{a posteriori} on data from coarse-grained simulations.

Finally we briefly described a fine-grained numerical model for the dynamics of a protein in solution. This model is formulated in a normal-model description, which includes both the rotational relaxation and the local energy barriers. The theory, called ``LE4PD'', predicts dynamics in agreement with experimental NMR relaxation, and does not require direct fitting of the experimental data.

\end{document}